\documentclass[useAMS,usenatbib]{mn2e}
\usepackage{times, subfigure}
\usepackage{import}
\usepackage{graphicx, subfigure}

\title[ SPIRE-FTS observations of IC342]{\it{Herschel}-SPIRE-Fourier
  Transform Spectroscopy of the nearby spiral galaxy
  IC342\thanks{Herschel is an ESA space observatory with science instruments provided by European-led Principal Investigator consortia and with important participation from NASA}}
\author[D.~Rigopoulou et al.]{D.~Rigopoulou,$^{1,2}$\thanks{E-mail:
d.rigopoulou1@physics.ox.ac.uk.} P.D.~Hurley$^{3}$,B.M.~Swinyard$^{2,4}$,
 J.~Virdee$^{1,2}$,
K.V.~Croxall$^{5}$,
\newauthor
R.H.B.~Hopwood$^{6,7}$,
T.~Lim$^{2}$,
 G.E.~Magdis$^{1}$, 
C.P.~Pearson$^{2}$,
E. Pellegrini$^{8}$,
\newauthor
 E.~Polehampton$^{2}$, 
J-D. Smith$^{8}$
 \\
$^{1}$Department of Astrophysics, University of Oxford, Keble Road, 
                Oxford, OX1 3RH, UK\\
$^{2}$RAL Space, Rutherford Appleton Laboratory, Chilton, Didcot OX11
0QX, UK\\
$^{3}$Astronomy Centre, Dept. of Physics \& Astronomy, University of
Sussex, Brighton BN1 9QH, UK\\
$^{4}$Dept. of Physics \& Astronomy, University College London, Gower
St, London, WC1E 6BT, UK\\
$^{5}$Department of Astronomy, Ohio State University, 140 West 18th
Avenue, Columbus, OH 43210-1173, USA\\
$^{6}$Physics Department, Imperial College London, South Kensington
Campus, London SW7 2AZ, UK\\
$^{7}$ Department of Physical Sciences, The Open University, Milton Keynes MK7 6AA, UK\\
 $^{8}$Department of Physics and Astronomy, University of Toledo, Toledo, OH 43606, USA
}

\date{Accepted 2013 June. Received 2013 June 7; in
  original form 2013 February 12}

\pagerange{\pageref{firstpage}--\pageref{lastpage}} \pubyear{2002}
\begin{document}
\maketitle

\label{firstpage}

\begin{abstract}

We present observations of the nearby spiral galaxy IC342 with the
Herschel Spectral and Photometric Imaging Receiver (SPIRE)  Fourier
Transform Spectrometer. The spectral range
afforded by SPIRE, 196-671 $\mu$m, allows us to access a number of 
$^{12}$CO lines  from J=4--3 to J=13--12 with the highest J
transitions observed for the first time. In addition we present
measurements of $^{13}$CO, [CI] and [NII]. 
We use a radiative transfer code coupled with Bayesian likelihood
analysis to model and constrain the 
temperature, density and column density of the gas.
We find two $^{12}CO$ components, one at 35 K and one at 400 K with 
CO column densities of 6.3$\times$10$^{17}$ cm$^{-2}$ and
0.4$\times$10$^{17}$ cm$^{-2}$ and CO gas masses of
1.26$\times$10$^{7}$ M$_{\odot}$
and 0.15$\times$10$^{7}$ M$_{\odot}$, for the cold and warm
components, respectively. 
 The inclusion of the high-J $^{12}$CO line observations, indicate the existence of a much warmer gas
component ($\sim$400 K) confirming  earlier findings from H$_{2}$
rotational line analysis from ISO and Spitzer. The mass of the warm
gas is 10\%  of the cold gas, but it likely dominates the
CO luminosity. In addition, we detect
strong emission from [NII] 205 $\mu$m and the 
$^{3}P_{1}\rightarrow^{3}P_{0}$ and $^{3}P_{2}\rightarrow ^{3}P_{1}$ [CI] lines at 370 and
608 $\mu$m, respectively. The measured $^{12}$CO line ratios can be
explained by Photon-dominated region (PDR) models although additional
heating by e.g. cosmic rays cannot be excluded. The measured [CI] line ratio together with
the derived [C] column density of 2.1$\times$10$^{17}$ cm$^{-2}$ and
the fact that [CI] is weaker than CO emission in IC342 suggests that 
[CI] likely arises in a thin layer on the outside of the  CO emitting
molecular clouds consistent with PDRs playing an important role.

\end{abstract}


\section{Introduction}

Far-infrared fine-structure lines of abundant elements such as carbon,
oxygen, nitrogen and sulphur either in their neutral or ionised state
contribute significantly to the gas cooling of the interstellar medium
(ISM, e.g. Hollenbach \& Tielens 1999). Far-infrared lines of ionised
atoms are useful probes of HII regions while the main cooling
of the neutral ISM is carried out by [CII] and [OI] (e.g. Malhotra et
al., 2001). In molecular gas, cooling is due
to [CI] and the carbon monoxide molecule CO. As potential
tracers of the gas cooling, submillimeter [CI] and CO lines are
expected to provide information on the gas heating rate, which is
dominated by the incident FUV radiation, mainly due to massive and
young stars. 

Neutral atomic carbon can be found in all
types of neutral clouds from diffuse to molecular. The ratio of the
two ground state fine-structure lines is a sensitive tracer of the
total gas pressure (e.g. Gerin \& Phillips 2000). Emission from the
two ground state fine-structure lines of atomic carbon is seen by COBE
throughout the Milky Way and makes a significant contribution to the
gas cooling. 
Despite the high abundance of atomic carbon [CI] in cool interstellar
media and the importance in controlling the overall thermal budget
only a handful of measurements of the ground state fine structure
lines at 370 and 608 $\mu$m have been achieved from the ground. The
first detection was reported by Buttenbach et al. (1992) in IC342. A
handful of galaxies have been detected since, including NGC 253
(Harrison et al. 1995), M82 (Stutzki et al. 1997), M83 (Petitpas \&
Wilson 1998), and M33 (Wilson 1997). 

The molecular CO transitions have been extensively studied
from the ground. However, the diagnostic power of the CO rotational
transitions has not been fully exploited since only the lowest
transitions are easily accessible with ground-based telescopes. In
recent years higher rotational transitions have been observed in a
handful of nearby (mostly) starburst galaxies (e.g. Papadopoulos et
al. 2010). The so called ``CO
Spectral Line Energy Distribution (SLED)'' 
 is used to probe the physical properties of the molecular
gas such as temperature and column density.

IC342 is a nearby (D=1.8 Mpc, 1''=8.7pc, McCall 1989) 
spiral galaxy. {\bf Because of its proximity, face-on grand spiral
appearance, enhanced star-forming activity in the central region
(e.g. Becklin et al. 1980)  and strong millimetre and submillimetre emission
IC342 has been a popular target for infrared and submillimetre
observations. With a far-infrared luminosity
1.25$\times$10$^{10}$L$_{\odot}$ (Dale et al. 2012), IC342 has been an early target of many
investigations of molecular gas and/or atomic far-infrared fine structure
lines.}  [CI] emission has already been detected from the ground
(Buttenbach et al. 1992) 
while a number of CO transitions have
also been observed (e.g. Bayet et al. 2004, 2006) allowing some 
constraints to be placed on the properties of the neutral and molecular ISM. 

In this paper we present new spectroscopic observations of
IC342 obtained using the SPIRE instrument on the Herschel Space
Observatory (Pilbratt
et al. 2010) covering the 194 to 671 $\mu$m regime. This spectral region is particularly
important as it allow us to access a number of high-J CO line transitions and
consequently investigate the properties of the molecular gas. In
particular, the current observations allow us to probe the peak of the CO SLED
and enable modeling of the physical properties of the molecular
ISM. In addition the detection of the two ground state [CI] and [NII]
lines allow us to investigate the conditions in the PDRs. The current
observations demonstrate the power of far-infrared and submillimetre
spectroscopy to probe the diffuse and ionised media in external
galaxies. The imminent availability of the Atacama Large
Millimetre Array (ALMA) will enable this kind of science in distant
galaxies. With a moderate FIR luminosity IC342 is representative of a
typical high-redshift galaxy and can serve as a template when
designing ALMA observations.

\section{Observations, Data Reduction and Results}
The present observations were taken as part of Herschel's Performance
Verification (PV) phase using SPIRE
(Griffin et al. 2010) as an imaging Fourier-Transform
Spectrometer (FTS). The SPIRE astronomical calibration methods and
accuracy have been presented in Swinyard et al (2010). SPIRE-FTS was
used in the high spectral resolution mode, sampling across a field of view of
2.6' in diameter.  The pointing of the observation was centered at R.A.=56.70322 deg. and
Dec=68.09614 deg and the total on-source integration time was 
9240 seconds. The SPIRE-FTS measures the Fourier transform of
the spectrum of a source using two detector arrays: SSW covering
the 194-313 $\mu$m and SLW covering 303--671 $\mu$m wavelength
bands simultaneously. The FWHM beamwidths of the SSW and SLW arrays
vary from 17'' at 194 $\mu$m to 42'' at 671 $\mu$m, respectively.
The size of the beams varies within this range in a complex fashion
due to the nature of the SPIRE detectors (Swinyard et al. 2010, Wu et
al. 2013). 
The FTS observations consisted of 132 seconds repetitions using single
pointing mode, sparse spatial sampling and high spectral resolution
(FWHM$\sim$0.048cm$^{−1}$).  The data were processed using the 
standard pipeline described in the Observers Manual (SPIRE Observers
Manual 2012) and Fulton et al. (2008)

The interferograms were cosmic ray,
temperature and time-domain phase corrected. The repetitions were then
averaged and Fourier transformed into the spectral domain. By taking
the inverse transform of the observed interferogram we can restore the
original source spectrum.
Although IC342 is a grand-design spiral, it has been found that a
significant fraction of the CO J=$1 \rightarrow 0$ and J=$2 \rightarrow$1 emission arises
from a $\approx$20''$\times$13'' central region (e.g. Eckart et al. 1990) which is
well matched to the size of the FTS SSW and SLW beams.

The (uncorrected) FTS spectrum of IC342 is shown in Figure 1 (right)
in red. Blue asterisks indicate the peak fluxes taken from the SPIRE photometric images
(these are quoted in fluxes$/$beam, Dale et al. 2012).
The prominent mismatch between SLW and SSW is the result of
two effects: the variations of the FTS spectral response with frequency and, of 
the way the beam couples to sources of varying
extent. The first effect has been modeled with a combination of
Hermite-Gauss (HG) polynomials, which follow the form of the expected
native feedhorn modes (Ferlet et al., in prep.). The efficiency with
which the beam couples to a given source has been estimated using
observations of Neptune and Uranus to establish the point source
response,  and a model of the flux from the Herschel telescope itself
to establish the fully extended response. IC342 fills the SSW beam
(low frequency) but
can be considered a point source in the SLW beam (high frequency).
Following the method described above and detailed in Fletcher et
al. (2012) and Wu et al. (2013, in prep) we infer the FWHM size of the
source to be $\sim$19 arcsec. The corrected FTS spectrum
(corresponding to a 19'' FWHM core) is shown in Figure 1 (black). 

\begin{figure}
 \centering
 \includegraphics[width=80mm]{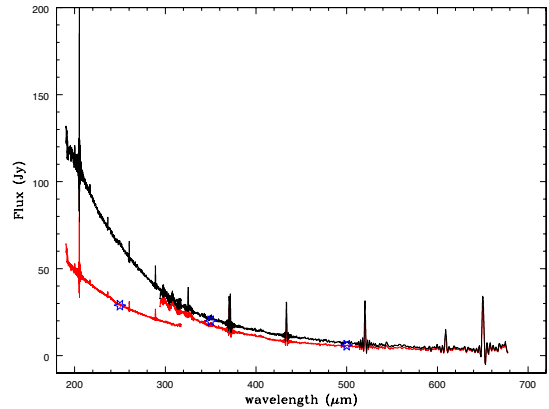} 
 \caption{The uncorrected spectrum (in red) showing a mismatch between
 SSW and SLW due to the difference in beam sizes and source
 extent. The corrected spectrum (assuming a 19 arcsec FWHM Gaussian
 distribution for the source size is shown in black. Blue asterisks
 denote peak fluxes (flux$/$beam) taken from the SPIRE images
 (priv. comm. R. Kennicutt)}
\end{figure}
We measured line fluxes from the calibrated unapodized spectrum
using our own IDL-based routines. In brief, from each of the SSW (SLW)
spectrum we first fit the underlying continuum which must be removed
before fitting the lines. After subtracting the continuum fit each line is fitted separately
using a sinc function with central frquency, line width, amplitude and residual
value(in most cases this equals zero since we have removed the
continuum) as free parameters. 
We also measured fluxes using another IDL based line fitting
program called SLIDE developed by
A. Rykala (priv. comm.). The integrated line fluxes derived from the
two independent methods agree very well. We note that the errors
reported in Table 1 are 1$\sigma$ uncertainties estimated from the
fitting method. 
Figure 2 shows the atomic, ionic and molecular lines identified in the
IC342 FTS spectrum. 
\begin{figure*}
 \centering
 \includegraphics[width=140mm]{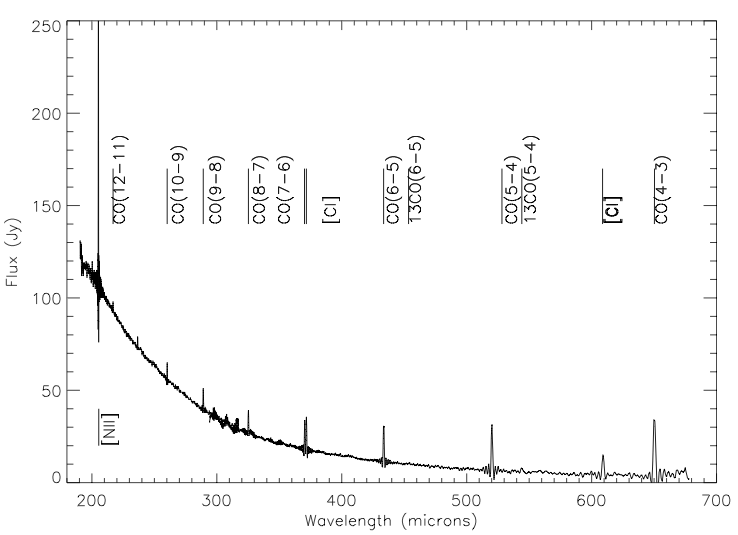} 
 \caption{The FTS spectrum of IC342 with the atomic, ionic and
   molecular lines identified.}
\end{figure*}


\begin{table}
 \centering
  \caption{Measured fluxes of detected emission lines}
  \begin{tabular}{@{}lcccc}  
  \hline
 Species&Transition     &Wavelength&F$_{\lambda}$&F$_{\nu}$d$\upsilon$ \\
 & & (rest, $\mu$m)  &(10 $^{-17}$ W m$^{-2}$)& (10$^{3}$ Jy km s$^{-1}$) \\
 \hline
$^{12}$CO& J=4--3&650.245&40.15$\pm$0.81$^{1}$&26.38 \\
$^{12}$CO&J=5--4&576.268&  35.62$\pm$0.75&18.73\\
$^{12}$CO &J=6--5&433.556&  31.23$\pm$0.69&13.68\\
$^{12}$CO& J=7--6&371.650&   24.23$\pm$0.87&9.10\\
$^{12}$CO& J=8--7&325.225&   19.27$\pm$0.82&6.33  \\
$^{12}$CO& J=9--8&289.118&   17.12$\pm$1.14&5.00 \\
$^{12}$CO& J=10--9&260.238&  12.52$\pm$0.73&3.30 \\
$^{12}$CO& J=11--10&236.611  &8.912$\pm$0.95& 2.13\\
$^{12}$CO& J=12--11&216.925   &6.342$\pm$0.82&1.39 \\
$^{12}$CO& J=13--12&200.271   &3.59$\pm$1.24& 0.72\\
\hline
 $^{12}$CO&J=1--0&2606.869&2.25$\pm$0.21$^{2}$&5.93 \\
$^{12}$CO&J=2--1&1303.434&8.85$\pm$0.76$^{2}$&11.66 \\
$^{12}$CO&J=3--2&868.956&22.51$\pm$1.19$^{2}$&19.77\\
\hline
$^{13}$CO& J=5-4&544.156&3.31$\pm$0.85&1.82\\
$^{13}$CO&J=6-5&453.494&1.18$\pm$0.40&0.54 \\
\hline
$[NII]$& $^{3}$P$_{1} \rightarrow ^{3}$P$_{0}$&205.226&117.65$\pm$8.2&54.64\\
$[CI]$ &$^{3} $P$_{1}\rightarrow ^{3}$P$_{0}$&370.466&21.70$\pm$0.8&8.124\\
$[CI] $&$^{3}$P$_{2}\rightarrow ^{3}$P$_{1}$&608.812&9.04$\pm$0.9&5.57\\
\hline
\end{tabular}
$^{1}$Quoted errors represent 1$\sigma$ errors from the line fitting
procedure and do not include (e.g.) instrumental uncertainties. \\
Values from Bayet et al (2006) and references therein, data
taken with beamsize 21.9''.
\end{table}

\section{Modeling the CO lines}


The properties of the molecular gas in IC342 have been studied through
low-J (J$<$6) CO emission lines by, amongst others,
Eckart et al. (1990), Harris et al. (1991) and Bayet et al. (2006). Extensive modeling of
the low-J CO
lines by Bayet et al. (2006) revealed gas densities of the order of a
few $\times$10$^{3}$ cm$^{-3}$ and gas (kinetic) temperatures around
40 K, for the low density, low temperature component. 
Through spatially resolved kinematics of the CO(6--5) line,  Harris et al (1991)
suggest that the warm gas is not originating from the same location as
the CO low-J lines.
With the  SPIRE-FTS we can now sample a large part of the $^{12}$CO-ladder
from J=4--3 to J=13-12. These high-J CO lines, inaccessible from the ground,
supplemented with detections of lower-J CO transitions allow us to
carry out a full investigation of the physical conditions 
of the molecular gas in IC342. 



For the present investigation of the physical conditions of the molecular
gas we used the non-LTE radiative transfer code RADEX (van der Tak et
al. 2007) to compute CO intensities (from J=1--0 to J=13--12) for a
large grid of temperatures T$_{kin}$, density n(H$_{2}$), column
density (N$_{\rm CO}$) and source size. We use the uniform expanding sphere
approximation and a 2.73 K blackbody to
represent the cosmic microwave background (CMB). As discussed by
Kamenetzky et al. (2012) the choice of background does not affect the
resulting kinetic temperature of the model component(s). The code
starts off with the optically thin case and generates level
populations. The process continues until a stable self-consistent
solution is found where the optical depth of the lines remain stable
from one iteration to the next. Using the code we have searched a
large grid of parameters in T$_{\rm kin} $: 10 -- 3000 K, n(H$_{2}$) :
10$^{2}$ -- 10$^{8}$ cm$^{-3}$, N$_{\rm CO}$ : 10$^{15}$--10$^{24}$
cm$^{-2}$ and, source size : 0--1000 arcsec$^{2}$. 
During the search procedure we reject those models where: 
(a) the optical depth of the low-J lines modeled is outside the range
-0.1 $<\tau<$100 (e.g. Van der Tak 2007, Kamenetzky et al. 2012) and (b) M$_{gas}$ becomes larger than
M$_{\rm dyn}$, the dynamical mass of the galaxy. We note that M$_{\rm gas}$ is defined as:
\begin{equation}
M_{\rm gas} = \Omega  D^{2}_{\rm A} N_{\rm CO} \times \frac{\mu m_{H_{2}}}{x_{\rm CO}} 
\end{equation}
\noindent
where, x$_{CO}$=3$\times$10$^{-4}$  is the relative CO$/$H$_{2}$
abundance, D$_{\rm A}$ is the
angular distance diameter  in cm$^{2}$, $\Omega$ is the
angular source size and $\Omega \times$ D$_{\rm A}^{2}$ is the source size.
The mean molecular weight, $\mu$ =1.4, is in units of m$_{H_{2}}$. The
source size remains the same for all transitions. We finally
assume a line width of 54 km s$^{-1}$ (Bayet et al 2006).

The parameter space is searched using the nested sampling routine
MULTINEST (Feroz et al. 2008). In brief, MULTINEST is a Bayesian
inference tool for model selection and parameter estimation. It is
based on the Monte Carlo technique of nested sampling (Skilling 2004),
which can evaluate the Bayesian evidence (useful for model selection)
and sample from posterior distributions with (often an unknown number
of) multiple modes and$/$or degeneracies between parameters. The
posterior distribution P$_{r}(M|x)$ gives the probability of the model
parameters (M) given a set of measurements x. Using Bayes theorem, the
posterior can be expressed as:

\begin{equation}
P_{r}(M|x) = \frac{P_{r}(M)P_{r}(x|M)}{P_{r}(x)}
\end{equation}
\noindent
where P$_{r}(M)$ is the prior probability that a set of parameters is
either physical or unphysical and is set to the grid ranges described
above.  P$_{r}(M|x)$  is the likelihood of reproducing the observational
measurements with a RADEX SLED, given a specific set of model
parameters. We use the standard Gaussian likelihood measure to
calculate the likelihood. P$_{r}(x)$ is the normalisation parameter or
Bayesian Evidence which is used for model selection. In order to find
the posterior distribution for one parameter, e.g. T$_{\rm kin}$, we need to
marginalise over all other parameters to find the probability
P(T$_{\rm kin}$).

\section{Results and Discussion}

We use RADEX and the nested sampling routine MULTINEST, described
above, to model the CO line intensities.  All the CO lines reported in
Table 1
are used in the
analysis supplemented with low-J CO lines reported in the
literature. For the total uncertainty, we take the
1$\sigma$ statistical uncertainty in the total integrated intensity from the
line fitting procedure and add 10\% calibration error in quadrature. 
We first consider a one-component model (assuming a uniform temperature for all
transitions). The highest likelihood model is shown in Figure 3.
The resulting `posterior' distributions for each parameter
marginalised over the other parameters used in the model are shown in
Figure 4. In Table 2 we report the mean, standard deviation and
maximum likelihood (ML) values for the T$_{\rm kin}$, $n(H_{\rm 2})$,
N($_{\rm CO}$) and $\Omega$. 

\begin{figure}
 \includegraphics[width=1.0\hsize, angle=0]{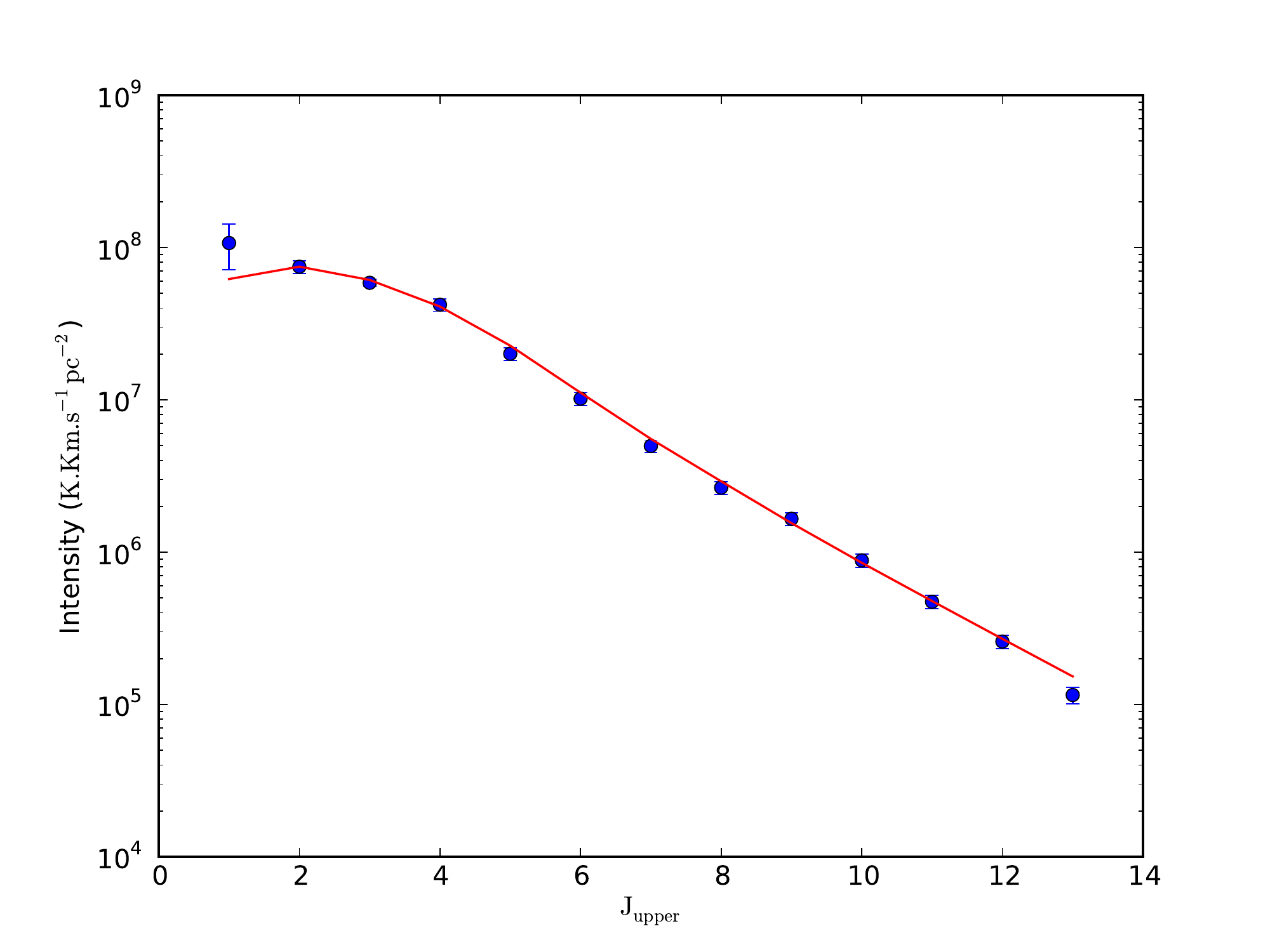}
\caption{The highest likelihood model (red line) is compared to our
  data (blue points) and associated errors.}
\end{figure}
\begin{figure}
 \includegraphics[width=0.9\hsize, angle=0]{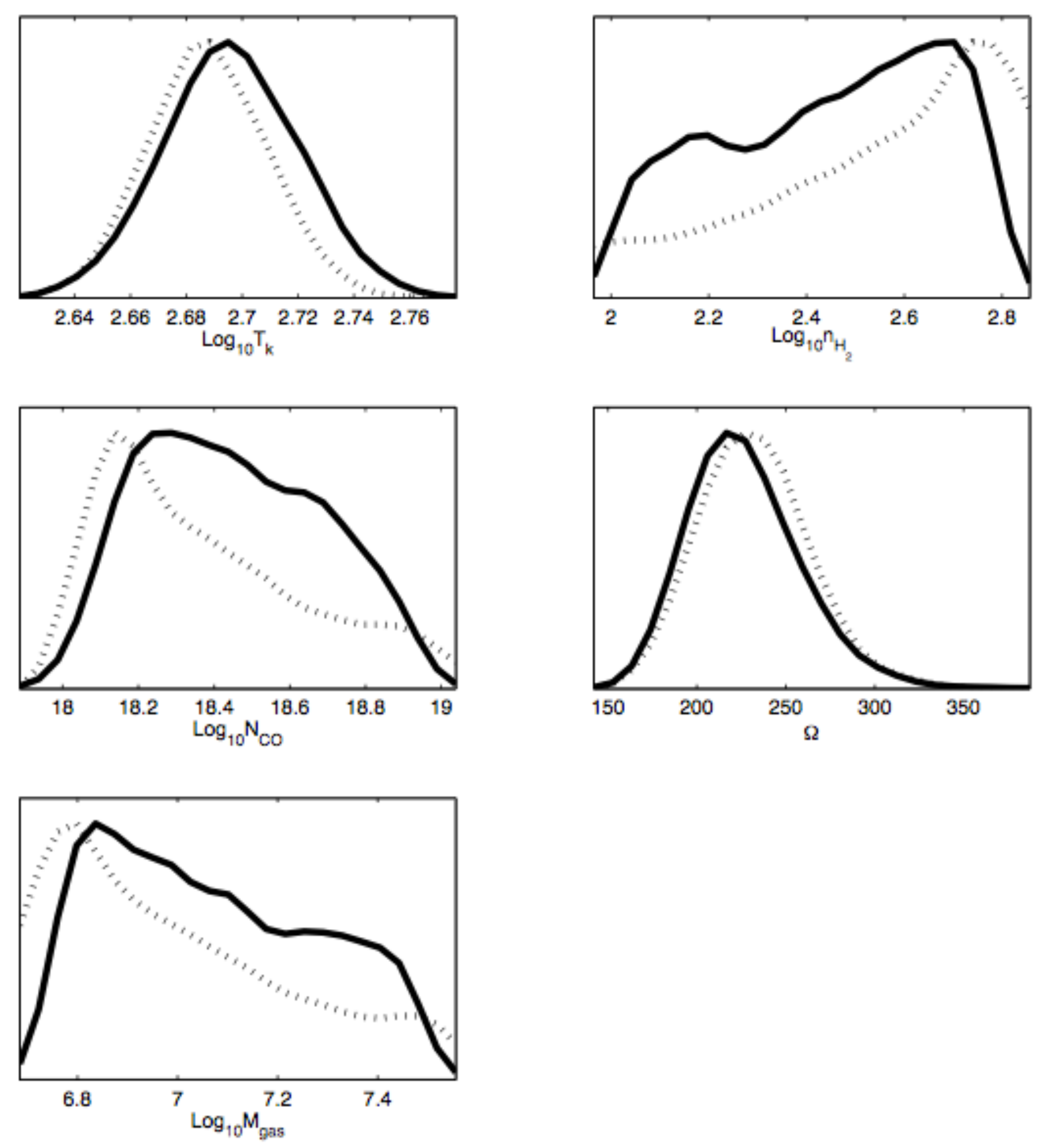}
\caption{Maximum Likelihood distributions of kinetic temperature,
   H$_{2}$ density, CO column density, M$_{gas}$ and $\Omega$ source extent. The dotted lines
 represents the mean distributions of the values. }
\end{figure}

\begin{table}
 \centering
\caption{Parameters of the single component model}
  \begin{tabular}{@{}lccccc}  
  \hline
  &T$_{\rm kin}$&n(H$_{2}$)&N$_{\rm CO}$&M$_{\rm gas}$&$\Omega$\\
 & K & cm$^{-3}$&$\times$10$^{18}$ cm$^{-2}$ &$\times$10$^{7}$M$_{\odot}$&arcsec\\
\hline
mean&398&251 &2.5&1.15&227\\
std dev&1.04&1.65&1.69&1.62&40\\
ML&398&549&1.25&0.63&224\\
\hline
\end{tabular}
\end{table}

The deviation of the low J (J$\leq$3)
transitions from our highest likelihood model 
indicates the presence of a second (possibly colder) gas component primarily responsible  
for the low-J lines.  We thus consider a two-component model to fit the
CO-SLED following the same procedure as before.
In our two component model we allow the three RADEX parameters
$\mathrm{T}_{kin}$, $\mathrm{n(H_{2})}$ and $\mathrm{N}_{CO}$ to vary
independently for both components, giving us 6 free parameters to
search over. We restrict the temperatures of the two components to be
in the range 10$<T_{k}<$200K and 200$<T_{k}<$3000, thus, defining a warm and cold
component. We fix the source size to the size of the
FTS beam ($2140 arcsec^{2}$) as the cold gas is likely to fill the entire beam
and, place an upper limit to the dynamical mass of the combined gas mass from
both components.

Unlike previous studies that model the CO-SLED of galaxies
(e.g. Kamenetzky et al. 2012) we do not restrict the components to
any particular subset of CO lines. That is, both components are
combined to give the overall line luminosity.  Since there are few constraints
on the actual source size of the emitting region from the different
CO lines, we also fix the source size of both components to the size
of the FTS beam (2140arcsec$^{2}$). In order to ensure that our
results are not significantly biased by fixing the source size we have
repeated our models allowing the source size to vary. 
Such runs have
indicated that there exist a degeneracy between source size and column
density which is rather difficult to break. As a result we decided to
fix the source size so that we can have an independent estimate for
the CO column density of the two components without concerns about
possible degenerate values. 

In Figure 5 we show  the highest likelihood two-component model
that provides the best fit to the available $^{12}$CO data. The low-J lines originate in the
cold gas component (35 K) while the warmer gas component (398
K) is responsible for the higher-J lines.
The marginalised plots for each
parameter with the mean likelihood shown as dotted line are presented
in Figure 6. In Table 3 we report the detailed
values of the physical characteristics of the warm and cold gas. 
\begin{figure}
 \includegraphics[width=1.0\hsize, angle=0]{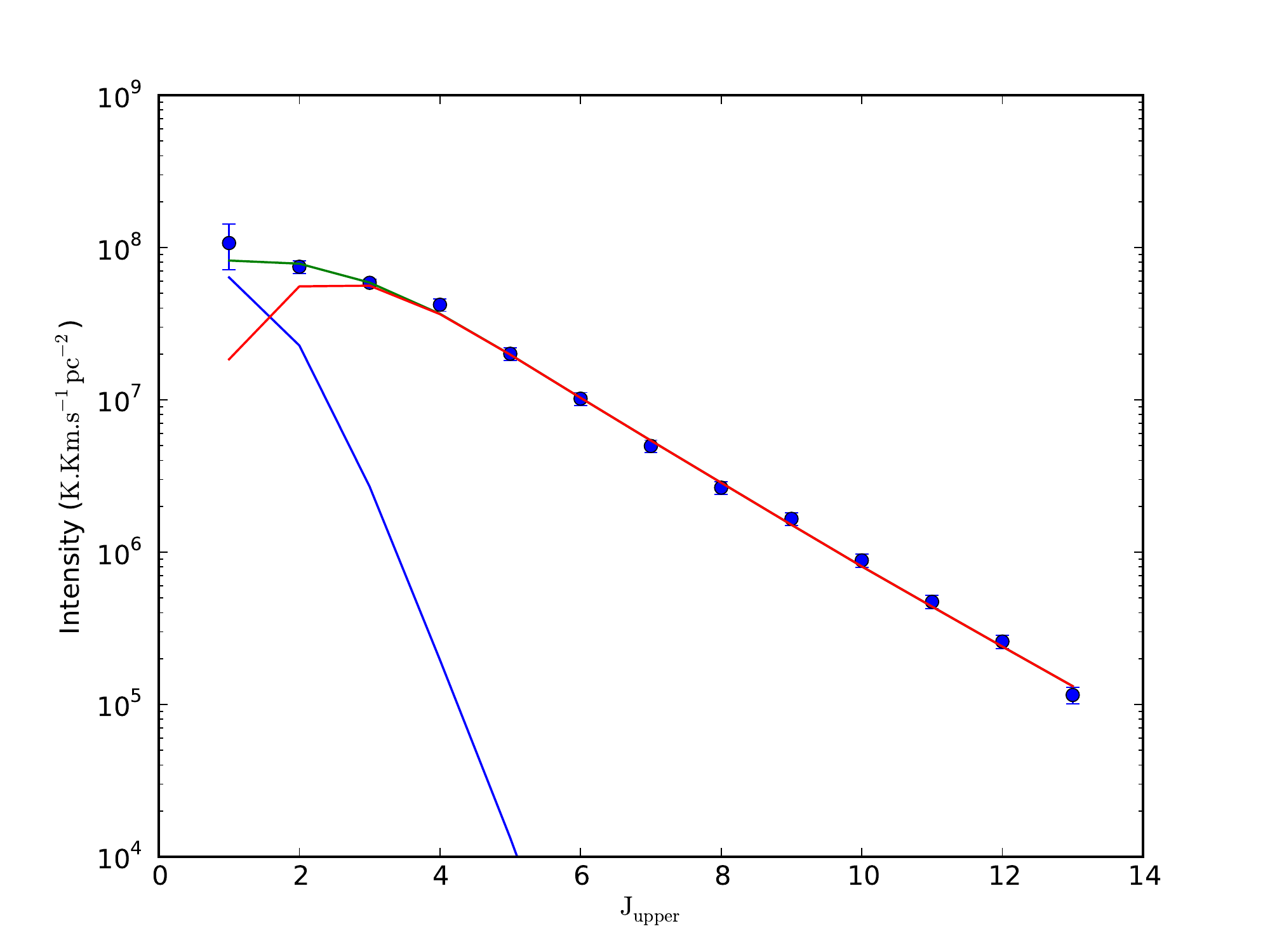} 
 \caption{ Maximum Likelihood Analysis  of the $^{12}$CO Spectral Energy
   Distribution. The CO measurments and associated error bars are
   shown as blue filled circles. The blue line represents the cold
   component while the red line represents the warm component. The
   total of the two components is the green line.}
\end{figure}

\begin{figure}
 \includegraphics[width=1.0\hsize, angle=0]{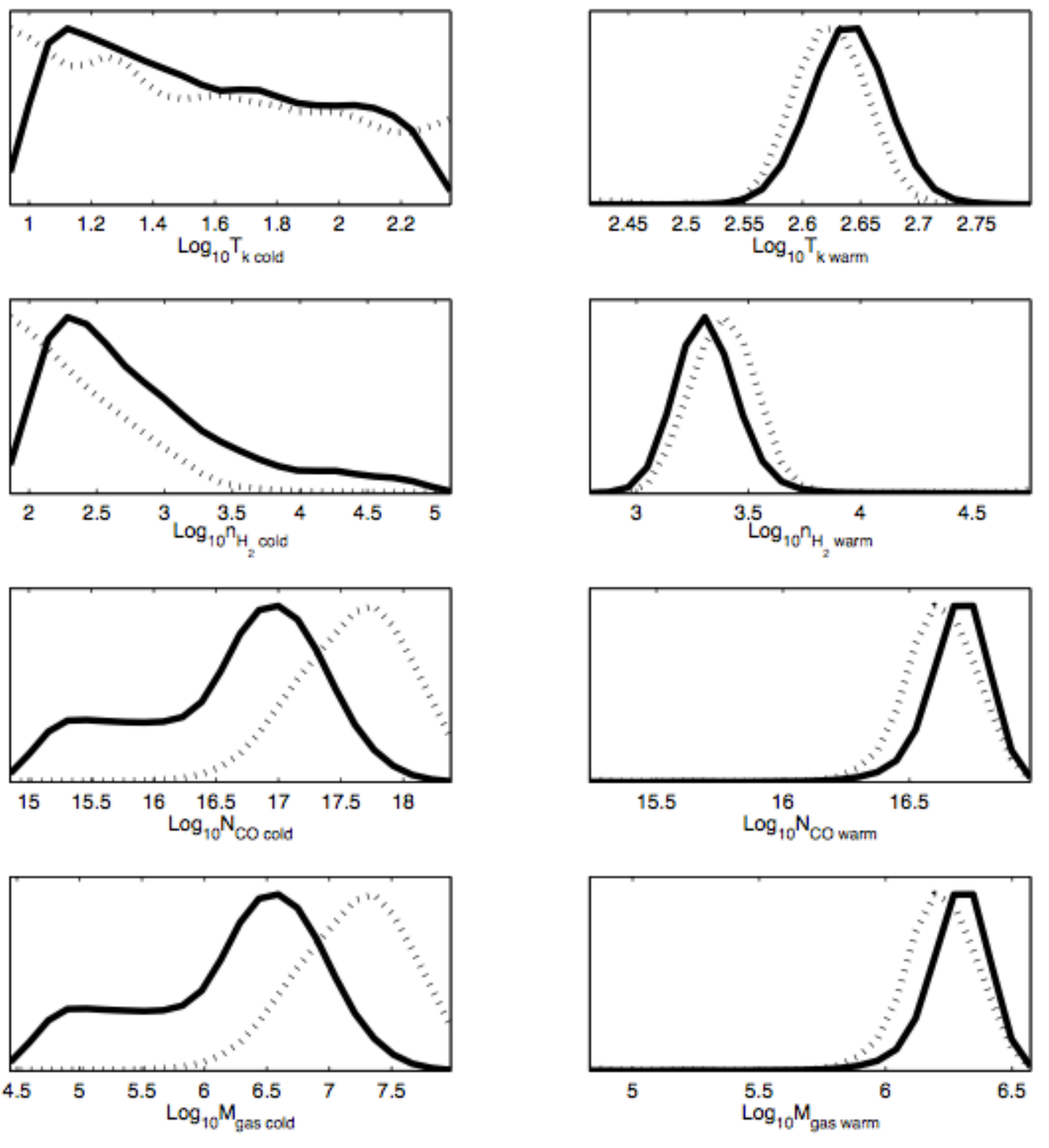} 
 \caption{ Posterior distributions of kinetic temperature,
   H$_{2}$ density, CO column density and M$_{gas}$ for the cold (left
 panel) and the warm (right panel) components. The dotted line
 represents the mean distributions of the values.}
\end{figure}
In Figure 7 we show the 2-d marginalised contour plots for the three main
parameters, T$_{kin}$, $\eta_{H_{2}}$ and N$_{CO}$. The plots
(1 and 2$\sigma$ contours) show the range of parameters for the two
temperature components (blue contours for the cold and red contours
for the warm component). The range of temperatures for the cold
component is much wider than for the warm component. Two reasons
are likely contributing to this effect. First, MULTINEST is very
sensitive to errors (of the CO line fluxes) and the error reported
for the CO(1--0) transition is driving the fit. Second, for the two
component model, we have fixed the apparent source size to the 
size of the FTS beam. It is very likely that the cold gas may
originate in a more spatially extended area than gas at warmer temperatures.
The equivalent parameters (T$_{kin}$,
versus $\eta_{H_{2}}$ and N$_{CO}$) for the warm component show a 
much narrower variation. 

Following equation 1, in Table 3 we report the mean, standard deviation and MLM values for
the gas mass for the each of the two
components considered. For the the warm component the
MLM warm gas mass value is 1.6$\times$10$^{6}$ M$_{\odot}$ with a similar
mean value of 1.5$\times$10$^{6}$
M$_{\odot}$. Rigopoulou et al. (2002) reported the detection of
mid-infrared rotational H$_{2}$ transitions using ISO. Assuming an
ortho-to-para ratio of 3 the S(1)--S(2) ratio can be used to estimate
the temperature of the warm gas. For IC342 they report a warm gas
temperature of 365 K which is consistent with our value of 398 K
(taking into account differences in the beam sizes and
calibration). Based on their sample, Rigopoulou
et al. find that the warm gas mass (measured from H$_{2}$) accounts for 1 to 10\% of the total
gas mass (measured from CO) in starburst galaxies which is consistent with our findings. 

\begin{figure}
 \includegraphics[width=1.0\hsize, angle=0]{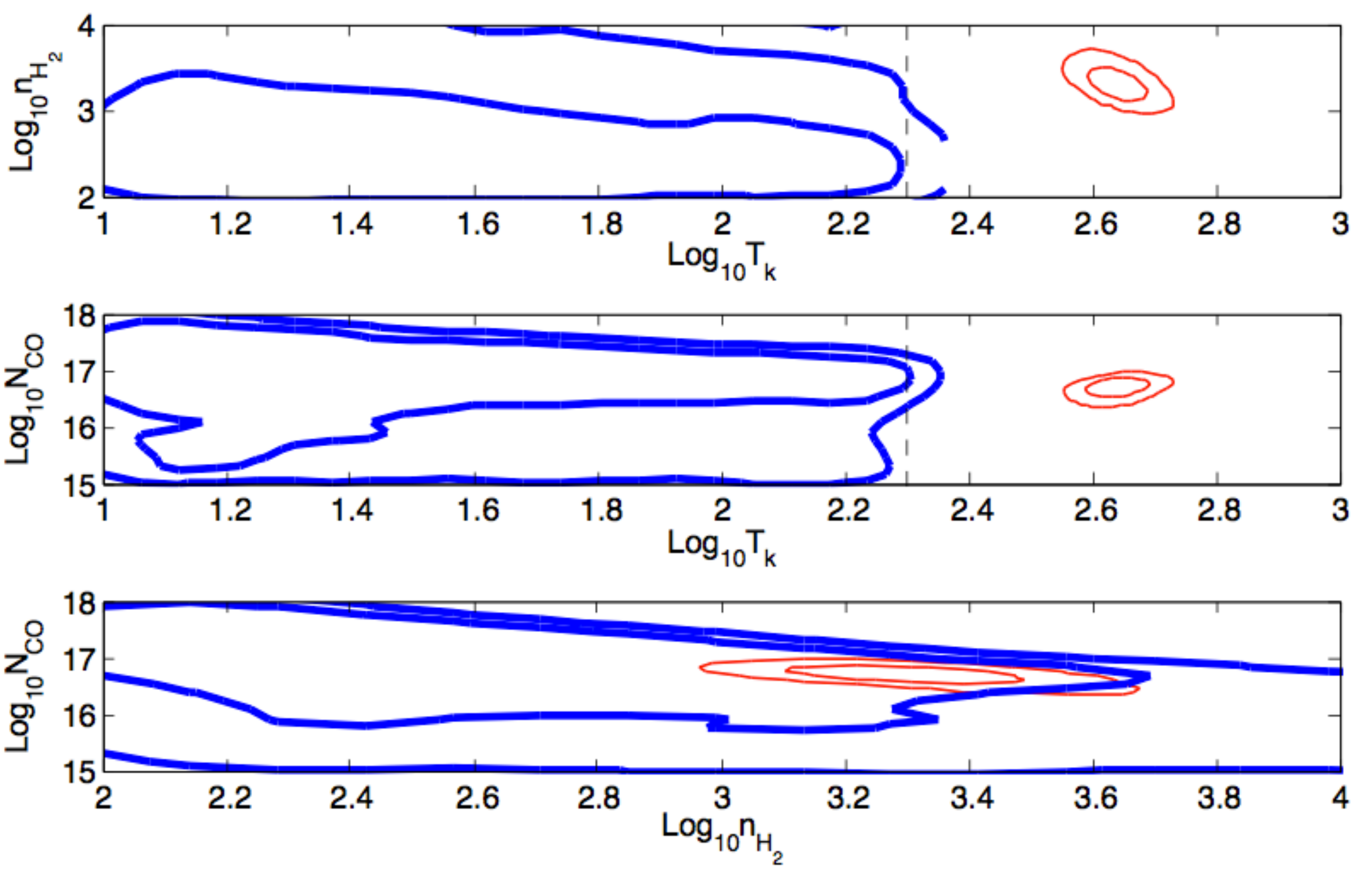} 
 \caption{One and two sigma contour levels for the three parameters
   $\eta_{H_{2}}$, N$_{CO}$ and T$_{kin}$ for the cold (blue) and the
   warm (red) component. }
\end{figure}

\begin{table}
 \centering
\caption{Parameters of two-component model}
  \begin{tabular}{@{}lccccc}  
  \hline
Cold Component& & & & \\
\hline 
 &T$_{\rm kin}$&n(H$_{2}$)&N$_{\rm CO}$&M$_{\rm gas}$\\
 & K & cm$^{-3}$&$\times$10$^{17}$ cm$^{-2}$ &$\times$10$^{7}$M$_{\odot}$\\
\hline
mean&35.5&630 &3.16&1.26\\
std dev&2.34&1.8 &1.76&1.73\\
ML&32.11&145&6.3&2.57\\
\hline
Warm Component& & & & \\
\hline
mean&398&2089&0.40&0.15\\
std dev&1.06&1.24&1.202&1.202\\
ML&407&2218&0.39&0.16\\
\hline
\end{tabular}
\end{table}

The ISM in IC342 has been studied in the past using ground based
measurements of (primarily low-J)  CO transitions. 
Using LVG calculations Eckart et al (1990) suggested T$_{\rm kin} >$ 20~K,
Israel \& Baas (2003) reported T$_{\rm kin}$ = 100 --150~K while 
Bayet et al. (2006) fit their data with a model of
40 K. All three models predict densities n(H$_{2}$) of the order of
2--3 $\times$10$^{3}$ cm$^{-3}$. The T$_{\rm kin}$ value we derive for the
cold component is in good agreement with the values reported by Bayet et
al. (2006) and Eckart et al . (1990) but lower than the value reported
by Israel \& Baas (2003). {\bf Our value also agrees with the T$\sim$53$\pm$1~K
and the  T$_{warm}$$\sim$443$\pm$130~K  derived by Mauersberger et
al. (2003) for the cold and warm gas components based on NH$_{3}$ observations.}
 The maximum likelihood value for the H$_{2}$
density of the cold temperature component predicted by our model is,
however, lower than those reported in the
literature. This is perhaps not surprising given that the cold gas
component is in fact expected to have a lower density and a higher
M$_{\rm gas}$. The warm gas component however, has a higher
n(H$_{2}$) density, and a lower column density N$_{\rm CO}$.
Based on the the lowest J $^{13}$CO lines, Meier et al (2000, 2001)
inferred gas temperatures of 10--20~K and densities of $\approx$10$^{3.5}$cm$^{-3}$.
It is thus likely that there are multiple components to the molecular
gas. Denser gas
with n(H$_{2}$) $>$10$^{6}$ is also present in IC342 as traced
through CS and  HC$_{3}$N lines (Aladro et al. 2011). This dense gas
is unlikely to contribute dominantly to the observed CO emission, instead it
is found in the central core of the molecular clouds in IC342 where
evidently (e.g. Martin et al. 2009) is it used to sustain high
star formation efficiency.

\section{The Origin of the Warm CO molecular Gas}

A number of mechanisms can heat the CO molecular
gas, including UV starlight in photodissociation
regions (PDRs),  X-ray heated gas (possibly associated with the
presence of an AGN) in X-ray dominated regions (XDRs), mechanical heating
(e.g. turbulence dissipation), powerful shocks and heating by cosmic rays. 
The absence of a strong AGN in IC342 together with the shape of the CO SLED
can probably rule out XDRs as the origin of the warm CO emission. In the
case of XDR dominance the CO SLED becomes flat at higher-J CO
transitions (e.g. Mrk 231, van der Werf et al. 2010) which is clearly
not the case in IC342.

A cosmic-ray ionisation rate of
$\sim$5$\times$10$^{-17}$ s$^{-1}$ (e.g. de Jong, Dalgarno and Boland 1980) is
sufficient to heat the gas to about $\sim$10 K. For an n(H$_{2}$)
density of 6$\times$10$^{2}$ cm$^{-3}$, CO column density of
3.16$\times$10$^{17}$ cm$^{-3}$ and an excitation temperature of 35 K
the cosmic-ray ionisation rate is $\approx$ 10$^{-15}$ s$^{-1}$ which
is about 20 times higher compared to the standard rate. As Eckart et
al. (1990) argue this is unlikely to be the case, especially in the
nuclear region of IC342 as it would require a very strong 5 GHz non
thermal radio component which is not observed. Therefore, although cosmic rays may
contribute towards heating low temperature $\sim$10 K gas it is
unlikely to be the dominant mechanism of CO heating. 

Recent$/$ongoing star-forming activity in IC342
(e.g. Meier \& Turner 2005, Boker et al. 1997) has resulted in an
increased number of OB stars and therefore, in a more intense UV
radiation field in this galaxy. The UV radiation heats up the surfaces
of the molecular clouds by means of photoelectric heating and heating
through far-UV pumping of H$_{2}$ (e.g. Tielens \& Hollenbach 1985).
PDRs form in the outer layer of the clouds which are responsible for
CO emission from warm$/$hot gas. Using the grid of PDR models
presented in Kaufman et al. (1999, 2006) 
we investigate predictions for the various $^{12}$CO lines. The rest
of this section deals with models of the $^{12}$CO lines although for
simplicity we refer to them as CO lines. The PDR models
cover a wide range in H$_{2}$ density (10--10$^{7}$ cm$^{-3}$)
and G$_{0}$, the 
incident UV flux. The parameter G$_{0}$, called the Habing interstellar
radiation field, is defined as FUV flux =1.3 $\times$10$^{-4}$
$\times$ G$_{0}$ erg cm$^{-2}$ s$^{-1}$ sr$^{-1}$.  The FUV flux is
related to the FIR flux via the relation FIR flux=2$\times$ FUV. 
We note, however, that in our investigations of the PDR models we do
not seek to determine the parameters that best fit the observed line
ratios. 
Instead, in addition to the relative CO line intensities, we use the physical
conditions determined from our LVG modeling as priors for PDR models.
In our search we do not only consider
the ML (or mean) value, instead, we make use of our likelihood
analysis and take into account a larger parameter grid (see Figure 5).
\begin{figure}
\centering
  \includegraphics[width=80mm]{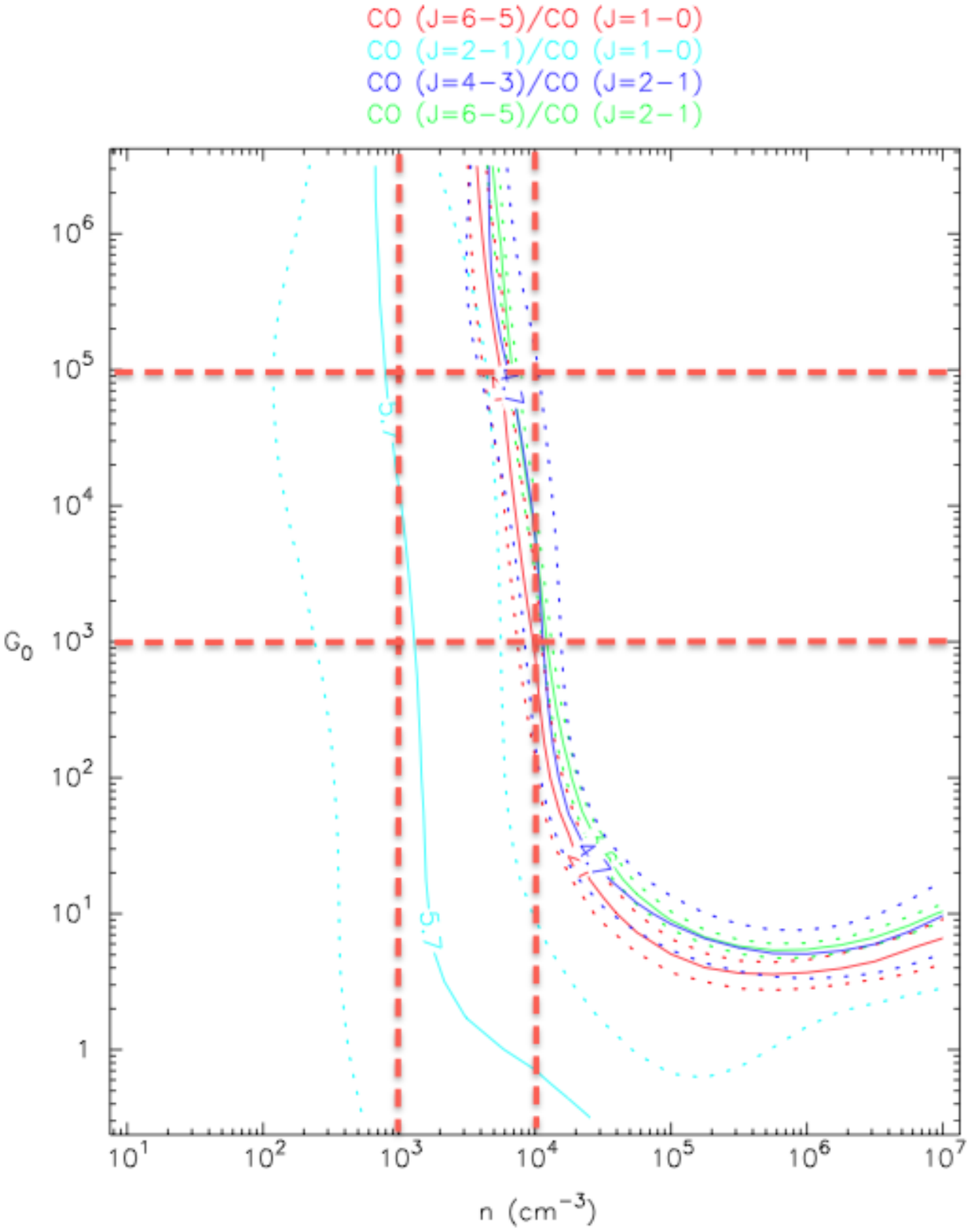}
\caption{\small  CO line ratios as a function of incident FUV flux (G$_{0}$ and
  and density $\eta$ for the PDR models. The vertical red dotted lines indicate the
  range of values for $\eta$ predicted by the RADEX models while the
  horizontal lines indicate the range of G$_{0}$ values required to
  match the observed CO line ratios. }
   \label{fig:sub:a}
\end{figure}
With the above caveats in mind, we have used the online PDR Toolbox
program \footnote {http://dustem.astro.umd.edu/pdrt/}
(Kaufman et al. 2006, Pound \& Wolfire 2008)
to investigate the range of PDR parameters that would best match our
CO line ratios. The PDR Toolbox program calculates the best
values of G$_{0}$ and cloud density $\eta$ for a given set of spectral
line intensities. A description of the PDR models used for these tools
can be found in Kaufman et al. (1999). For a given set of gas phase
elemental abundances and grain properties, each model is described by
a constant H nucleus density, n, and incident far-ultraviolet
intensity G$_{0}$. The models solve for the equilibrium chemistry, thermal
balance, and radiation transfer through a PDR layer. We used the line
intensities quoted in Table 1 supplemented by those from the
literature. The FTS lines have been measured assuming a FWHM of
19$^{''}$ as discussed in Section 2. Figure 8 shows
line ratios for various CO transitions. Assuming the range of
n(H$_{2}$) values determined from the likelihood analysis
(10$^{3}$--10$^{4}$ cm$^{-3}$) G$_{0}$ should be in the
range10$^{3} <$G$_{0}<$10$^{5}$ to match the observed CO line ratios.  In fact the
best-fit  PDR models predicts n(H$_{2}$) = 5.62$\times$10$^{3}$
cm$^{-3}$ and G$_{0}$=3.16$\times$10$^{5}$. 
Although at this stage a second mechanism (e.g. cosmic ray heating of
the coldest gas) contributing to the CO lines
cannot be excluded it is very likely that PDRs play an important,
perhaps dominant, mechanism for the CO
emission we have detected. Our measured line ratios can be matched
with PDR models with n(H$_{2}$) very close to the values found
by the maximum likelihood analysis. Based on the models presented we conclude that a
significant fraction of the CO emission in IC342 originates in PDRs.

\section{Far Infrared Atomic Fine Structure Lines}

As discussed in section 2 (and presented in Table 1) we detect three
fine structure lines, the two [CI] 370 and 608 $\mu$m and [NII] 205
$\mu$m lines. Prior to Herschel only a handful of [CI] extragalactic measurements
had been carried out from the ground. The [NII] 205 $\mu$m
line emission, in particular,  is very difficult to observe from the ground
because it lies near the long-wavelength cutoff for stressed Ga:Ge
photoconductors, strongly limiting the sensitivity of spectrometers in
this regime. Oberst et al. (2006) reported the detection of [NII] 205
$\mu$m towards the Carina galactic star forming region. With SPIRE-FTS
we now have additional [NII] 205 $\mu$m measurements in M82 (Panuzzo
et al. 2010, Kamenetzky et al. 2012), Arp220 (Rangwala et al. 2011).

The ratio of the [CI] lines can be used to estimate the physical
properties of the atomic [CI] gas that arises predominantly in the CO
dissociating regions. For IC342 we derive a I([CI]370$\mu$m)$/$I([CI]608$\mu$m) ratio of
1.45$\pm$0.3 which is close to the value  of 1.62 $\pm$0.2 using the
measurements reported for
M82 in Kamenetzky et
al. (2012) but higher than the value of 1.15$\pm$0.3 derived for Arp220
by Rangwala et al. (2011). We use the escape probability radiative
transfer models presented in Stutzki et al (1997) and their Figure 2,
to infer the [CI] column density N$_{\rm [CI]}$, T$_{\rm kin}$ and hydrogen
column density n(H$_{2}$). 
When expressed in line temperature units (K km s$^{-1}$) the 
[CI](J=2$\rightarrow$1)$/$[CI](J=1$\rightarrow$0) ratio for IC342
becomes 0.54. For this observed line ratio, a brightness temperature
of $\approx$ 15 K and assuming a line width of 54 km
s$^{-1}$ we estimate a column density N(C)=2.1$\times$10$^{17}$
cm$^{-2}$ and H$_{2}$ density of 5.4$\times$10$^{3}$
cm$^{-3}$. These values are indicative of an excitation temperature
40$<$T$_{\rm kin}$$<$60 K. The range of the inferred [CI] excitation temperature is
slightly higher than the one derived from cold (low-J) CO (mean value
of 36 K) although well within the uncertainties involved. It is, thus
likely that that [CI] may originate
in the same (cold) molecular gas. The column density ratio N$_{\rm C}
/$N$_{\rm CO}$ is 0.66 a value similar to that derived for M82 (e.g. Stutzki
et al. 1997) but lower than the value of 1 reported for Arp 220
by Rangwala et al. (2011). It has been suggested (e.g. Wilson et al. 1997
but see also discussion in Rangwala et al. 2011) that [CI] emission is
stronger for more luminous systems or those harboring intense
starbursts or AGN. 

Israel (2005) examined the [CI](2$\rightarrow$1)$/
^{12}CO(4-3)$ and [CI](2$\rightarrow$1)$/^{13}CO(2-1)$ line ratios for a sample of
quiescent, starburst and active galaxies. He found that in the
majority of galaxies the [CI](2$\rightarrow$1)$/^{13}CO(2-1)$
ratio is $\geq$2. Lower ratios are expected in high-UV environments
with high column densities where the majority of neutral C will be
locked up in CO. For IC342 the [CI](2$\rightarrow$1)$/^{13}CO(2-1)$ value of 1.2 
(Israel 2005) is smaller than the values found for the majority of
strong starbursts or luminous systems.  The line ratio together with
the derived C column density of 2.1$\times$10$^{17}$ cm$^{-2}$ and
the fact that [CI] is weaker than CO emission in IC342 could be
be indicative of 
[CI] arising in a thin layer on the outside of the  CO emitting
molecular clouds consistent with PDRs playing an important role.

The [NII]205 $\mu$m and [CII]158 $\mu$m lines have nearly
identical critical densities for excitation in ionised gas
regions. Their line ratio is thus insensitive to the hardness of the
stellar radiation field (since the photon energies required to ionize
each species to the next ionisation states are similar) and is only a
function of the [NII]205/[CII] abundance ratio. The ratio [NII]$/$[CII]
can, thus, give us an estimate of the fraction of [CII] arising in ionised
gas for a given value of the ionised gas density. The 
[NII] 122$/$205 $\mu$m line ratio can be used to probe the
density of the ionized gas. 

Using [CII] and [NII]122 $\mu$m measurements of IC342 from the ISO-LWS archive (but
also in Brauher et al. 2008 for [CII]) we infer a [NII] 122$/$205 $\mu$m line ratio
of 3.2 and a [CII]$/$[NII]205 ratio of 17$\pm$2.5.
We compare these
ratios to the model line intensity ratio as a
function of ionized gas density in Figure 2 of Oberst et
al. (2006). Their [NII] 122$/$205
model line ratios range from 2.5 to 4.3, therefore, our
estimated [NII] 122$/$205 $\mu$m line ratio of 3.2 indicates a density
of  $\sim$ 200 cm$^{-3}$. For this density, if both lines arise
from the ionized gas the expected [CII]$/$[NII] 205 ratio would have
been $\sim$3. Given our estimate of 17$\pm$2.5 we infer that 
the contributions from ionised gas to [CII] is 17-20\%
with the remaining originating in warm gas PDRs. This result
lends further support to the suggestion that PDRs must play an important role
in IC342.

\section{CO Ladder: an insight into the excitation of molecular gas of galaxies}

The most widely used method to investigate the properties of the
molecular gas of galaxies near and far, is through
measurements of CO transitions. 
For nearby galaxies, a suite of low-J CO transitions are readily accessible from the
ground and have been used to measure the excitation of the molecular gas (e.g. Boone et al. 2011). For
high-redshift galaxies however, we can only obtain a handful of CO transitions,
depending on redshift (e.g. Solomon \& Vanden Bout 2005, Wagg et
al. 2010), hence the resulting CO-SLEDs are
very sparsely sampled. The situation is likely to change with the
availability of the full ALMA array.

With the SPIRE-FTS we can now measure  a number of high-J CO
transitions (SPIRE-FTS can access up to J=13 although higher 
CO transitions up to J=20 can be observed with PACS) and when
combined with ground based measurements we can determine the CO-SLED
to much greater accuracy. As Kamenetsky et al (2012) point out, the diagnostic tools to model
the high-J CO lines are still under development, however, we have
already began to exploit the new information available. Extremely high
CO line excitation implies the presence of an AGN and strong heating
by XDRs or shocks. In star forming regions PDRs play a crucial role by heating the dust efficiently. 
The CO-SLED of Mrk 231  has been modelled as the result of
contributions from PDRs and XDRs, the latter to account for the
emission of the high-J lines (van der Werf et al. 2010).  In lower
luminosity AGN signatures of XDR heating are present close to the
nucleus on small scales while in the starburst regions CO excitation
originates in by PDRs (e.g Spinoglio et al. 2012). In Starbursts, such
as M82, detailed analysis of the CO-SLED revealed that although PDRs
could play a role they cannot provide high enough densities to match
the observed CO line emission. Kamenetsky et al. (2012) have therefore
concluded that mechanical energy (turbulent motions, shocks, cosmic
ray heating) is likely to play a large role in the heating of the
galaxy. A similar conclusion has been reached by Rangwala et
al. (2011) for Arp220. 

In the case of IC342, we have shown that the CO-SLED can be adequately
explained with a combination of a cold (T$\sim$35 K), low density
component to explain the low-J transitions (with most contribution
from J=1-0) and a warmer component (T$\sim$400 K) necessary to account for the high-J
transitions. {\bf There is convincing evidence that CO excitation in this quiescent star-forming galaxy 
can be provided by PDRs without invoking a need for extra heating
mechanisms. Thus, IC342 may be used as a template for explaining the
CO-SLED of high-redshift main-sequence galaxies} that do not appear
to be merger-driven (e.g. Elbaz et al. 2011).

\section{Conclusions}

We have presented spectroscopic observations of IC342 covering the
wavelength range 194-671 $\mu$m carried out with the SPIRE-FTS on
board Herschel. We have detected a number of $^{12}$CO, $^{13}$CO
molecular, [CI] atomic and [NII] ionic lines. We have used the
radiative transfer code RADEX coupled with MULTINEST to
model the $^{12}$CO SLED: the modeling procedure has revealed two gas components, a cold temperature
component T$\sim$35 K and, a much warmer component with T$\sim$400~K. 
Our new SPIRE-FTS data and in particular the high-J $^{12}$CO
lines have allowed us to constrain the physical properties of the warm gas component.
Based on the observed CO line ratios and published models of PDRs we
argue that the CO emission originates in PDRs although additional
contributions from e.g. cosmic ray heating cannot be excluded. 
Using the observed atomic [CI] line ratio we discuss the origin of
the [CI] and suggest that it arises in a thin layer on the outside of the  CO emitting
molecular clouds. This claim is further supported by the fact that the
[CI] emission in IC342 is much weaker than the CO(4--3). 
Using [CII] 158 $\mu$m measurements from ISO (Brauher et al. 2008) and
the [NII] 205 $\mu$measurement presented here we infer that up to 70\%
of [CII] arises in warm PDRs on the surface of molecular gas
clouds. 
Such detailed studies of the properties of the atomic, molecular and
ionised gas in nearby galaxies have only recently become available
thanks to the capabilities of Herschel. 
These studies are however,
very important since they provide templates for understanding the
physics of the ISM of high redshift bright submillimetre galaxies that
are currently impossible to study at the same level of detail. ALMA
however, in its full potential will be able to access the CO-ladder
of high redshift galaxies and extend our understanding of the
properties of the ISM in extreme environments.

\section{Acknowledgments}
GEM acknowledges support from the John Fell Oxford University Press
(OUP) Research Fund. DAR acknowledges useful discussions with Rob
Kennicutt, Daniela Calzetti and Chad Engelbracht and, in particular,
the availability of SPIRE data for IC342 ahead of publication.
SPIRE has been developed by a consortium of institutes led by
Cardiff Univ. (UK) and including: Univ. Leth- bridge (Canada); NAOC
(China); CEA, LAM (France); IFSI, Univ. Padua (Italy); IAC (Spain);
Stockholm Observatory (Sweden); Imperial College London, RAL,
UCL-MSSL, UKATC, Univ. Sussex (UK); and Caltech, JPL, NHSC,
Univ. Colorado (USA). This development has been supported by na-
tional funding agencies: CSA (Canada); NAOC (China); CEA, CNES, CNRS
(France); ASI (Italy); MCINN (Spain); SNSB (Sweden); STFC, UKSA (UK);
and NASA (USA).


\begin{thebibliography}{}
\bibitem[\protect\citeauthoryear{Aladro}{(2011}]{al} Aladro, R.,
  Martin-Pintado, J., Martin, S., Mauersberger, R.,
 Bayet, E., 2011, A\&A, 525, 89
\bibitem[\protect\citeauthoryear{Bayet}{2004}]{b4} Bayet, E., Gerin,
  M., Phillips, T. G., Contursi, A., 2004,
A\&A, 427, 45
\bibitem[\protect\citeauthoryear{Bayet}{2006}]{b6} Bayet, E., Gerin,
  M., Phillips, T. G., Contursi, A., 2006, A\&A, 460, 467
\bibitem[\protect\citeauthoryear{Becklin}{1980}]{b1} Becklin, E., et
  al., 1980, ApJ, 236, 441
\bibitem[\protect\citeauthoryear{Boker}{1997}]{b5} Boeker, T.,
  Forster-Schreiber, N. M., Genzel, R., 1997, AJ, 114, 1883
\bibitem[\protect\citeauthoryear{Boone}{2010}]{b11} Boone, F., et al.,
  2011, A\&A 525,18
\bibitem[\protect\citeauthoryear{Brauher}{2008}]{b08} Brauher, J.R.,
  Dale, D.A., Helou, G., 2008, ApJS, 178, 280 
\bibitem[\protect\citeauthoryear{Dale et al}{2012}]{d12} Dale, D., et
  al., 2012, ApJ 745, 95
\bibitem[\protect\citeauthoryear{deJongDalgarno}{1980}]{djdb} de Jong, T.,
  Boland, W., Dalgarno, A., 1980, A\&A, 91, 68 
\bibitem[\protect\citeauthoryear{Eckart}{1990}]{eck1} Eckart, A.,
Downes, D., Genzel, R., Harris, A. I., Jaffe, D. T., Wild, W., 1990,
ApJ, 348, 434
\bibitem[\protect\citeauthoryear{Elbaz}{2011}]{elb1} Elbaz, D., et
  al., 2011, A\&A, 533,119
\bibitem[\protect\citeauthoryear{Feroz}{2008}]{f2} Feroz, F., Hobson,
  M.P., 2008, MNRAS, 384, 449
\bibitem[\protect\citeauthoryear{Fletcher}{2012}]{f12} Fletcher, L.N., 
Swinyard, B., Salji, C., et al., 2012, A\&A, 539, 44 
\bibitem[\protect\citeauthoryear{Fulton}{2008}]{fu1} Fulton, T. R., 
Naylor, D. A., Baluteau, J.-P., et al. 2008, in SPIE Conf. Ser., 7010 
\bibitem[\protect\citeauthoryear{Gerin Phillips}{2000}]{g1} Gerin, M.,
  Phillips, T., 2000, ApJ, 537, 644
\bibitem[\protect\citeauthoryear{Griffin}{2010}]{g2} Griffin, M.J., et
  al. 2010, A\&A, 518, 3
\bibitem[\protect\citeauthoryear{HollenbachTielens}{1999}]{t1} Hollenbach,D.J.,
  Tielens, A.G.G.M., 1999, Rev. of Modern Physics, Vol. 71, 173 
\bibitem[\protect\citeauthoryear{IsraelBaas}{2003}]{ib} Israel,F.P.,
  Baas, F., 2003, A\&A, 404, 495 
\bibitem[\protect\citeauthoryear{Israel}{2005}]{ir1} Israel, F.P.,
  2005, A\&A, 438, 855
\bibitem[\protect\citeauthoryear{Kamenetzky}{2012}]{k3} Kamenetzky, J.,
et al.,   2012, APJ, 753, 70
\bibitem[\protect\citeauthoryear{Kaufman}{2006}]{k16} Kaufman, M., J., 
Wolfire, M. G., Hollenbach, D. J., 2006, ApJ, 644, 283
\bibitem[\protect\citeauthoryear{Malhotra}{2001}]{m1} Malhotra, S., 
Kaufman, M. J., Hollenbach, D., Helou, G., et al., 2001, ApJ, 561, 766 
\bibitem[\protect\citeauthoryear{Martin}{2009}]{m2} Martin S.,
  Martin-Pintado, J., Mauersberger, R., 2009, ApJ, 694, 610
\bibitem[\protect\citeauthoryear{MeierTurner}{2005}]{mt} Meier, D.S.,
  Turner, J.L., 2005, ApJ, 618, 259
\bibitem[\protect\citeauthoryear{Meier2}{2001}]{mei} Meier, D. S,
  Turner, J.L., Hurt, R.L., 2001, ApJ, 531, 200
\bibitem[\protect\citeauthoryear{Mauers}{2003}]{mau} Mauersberger, R.,
 Henkel, C., Weiß, A., Peck, A.B., Hagiwara, Y., 2003, A\&A, 403, 516
\bibitem[\protect\citeauthoryear{Oberst}{2006}]{o1} Oberst, T. E.,
  Parshley, S. C., Stacey, G. J., Nikola, T., et al., 2006, ApJ, 652, 125
\bibitem[\protect\citeauthoryear{Panuzzo}{2010}]{p3} Panuzzo, P., et
  al., 2010, A\&A, 518, 37
\bibitem[\protect\citeauthoryear{Papadopoulos}{2010}]{p1} Papadopoulos,
  P.P., van der Werf, P., Isaak, K., Xilouris, E. M., 2010, ApJ, 715, 775
\bibitem[\protect\citeauthoryear{PetipasWilson}{1998}]{pw1} Petitpas, G.,
  Wilson, C., 1998, APJ, 503, 219
\bibitem[\protect\citeauthoryear{Pilbratt}{2010}]{p1} Pilbratt, G., et
  al., 2010, A\&A, 518, 1
\bibitem[\protect\citeauthoryear{PoundWolf}{2008}]{p13} Pound, M.W.,
  Wolfire, M.G., 2008, ADAS XVII, 394, 654 
\bibitem[\protect\citeauthoryear{Rangwala}{2011}]{r3} Rangwala, N., 
Maloney, P. R., Glenn, J., Wilson, C. D.,  et al., 2011, ApJ, 743, 94
\bibitem[\protect\citeauthoryear{Rigopoulou}{2002}]{R3} Rigopoulou,D., 
Kunze, D., Lutz, D., Genzel, R., Moorwood, A. F. M., 2002, A\&A, 389, 374
\bibitem[\protect\citeauthoryear{Skilling}{2004}]{sk1} Skilling, J.,
  2004, AIPC, 735, 395
\bibitem[\protect\citeauthoryear{Spinoglio}{2012}]{sp3} Spinoglio, L., 
Pereira-Santaella, M., Busquet, G., et al., 2012, ApJ, 758, 108
\bibitem[\protect\citeauthoryear{SPIRE}{2011}]{SP111} 
SPIRE Observers Manual 2011, HERSCHEL-HSC-DOC-0798, v2.4
edn., Herschel Science Centre, $http://herschel.esac.esa.int/
  Documentation.shtml$
\bibitem[\protect\citeauthoryear{Stutzki}{1997}]{s1} Stutzki, J., et
  al.,  1997, ApJ 477, 33
\bibitem[\protect\citeauthoryear{Swinyard}{2010}]{s1} Swinyard, B., et
  al., 2010, A\&A 518, 4
\bibitem[\protect\citeauthoryear{vandenBout}{2005}]{vdb} Solomon, P.M.,
  Vanden Bout, P.A., 2005, ARA\&A, 43, 67
\bibitem[\protect\citeauthoryear{van der Werf}{2010}]{vdw} van der Werf,
  P.P., Isaak, K. G., Meijerink, R., Spaans, M., 2010, A\&A 518, 42
\bibitem[\protect\citeauthoryear{van der Tak}{2007}]{vdt1} van der Tak,
  F.,F. S.,  Black, J. H., Schoier, F. L., Jansen, D. J., van
  Dishoeck, E. F.,2007, A\&A 468, 627
\bibitem[\protect\citeauthoryear{Wang}{2010}]{wg} Wagg, J., et al.,
  2010, A\&A 519, 1
\bibitem[\protect\citeauthoryear{Wilson}{1997}]{w1} Wilson, C.D.,
  1997, ApJ, 487, 49
\bibitem[\protect\citeauthoryear{Young}{1986}]{y1} Young, J.S., et
  al., 1986, ApJ, 311, L17
\end{thebibliography}
\end{document}